\documentclass[aps,prb,superscriptaddress,showpacs,twocolumn,10pt,letterpaper]{revtex4-1}

\usepackage{hyperref}
\usepackage{graphicx}
\usepackage[english]{babel}
\usepackage[latin1]{inputenc}
\usepackage{amsmath}
\usepackage{amsfonts}
\usepackage[amssymb]{SIunits}
\usepackage{SIunits}
\usepackage{xfrac}
\usepackage{mathtools}

\def\BiSe{Bi$_2$Se$_3$}

\begin{document}

\title{Influence of Topological Edge States on the Properties of Al/Bi$_2$Se$_3$/Al Hybrid Josephson Devices}%

\author{L. Galletti}%
\email[L. Galletti: ]{luca.galletti@unina.it}
\affiliation{Dipartimento di Scienze Fisiche, Università degli Studi di Napoli Federico II, I-80126 Napoli, Italy}
\affiliation{CNR-SPIN Napoli, Complesso Universitario di Monte Sant'Angelo, 80126 Napoli, Italy}
\author{S. Charpentier}%
\affiliation{Quantum Device Physics Laboratory, Department of Microtechnology and Nanoscience, Chalmers University of Technology, SE-41296 Göteborg, Sweden}
\author{M. Iavarone}%
\affiliation{Department of Physics, Temple University, Philadelphia, Pennsylvania 19122, USA}
\author{P. Lucignano}
\affiliation{Dipartimento di Scienze Fisiche, Università degli Studi di Napoli Federico II, I-80126 Napoli, Italy}
\affiliation{CNR-SPIN Napoli, Complesso Universitario di Monte Sant'Angelo, 80126 Napoli, Italy}
\author{D. Massarotti}%
\affiliation{Dipartimento di Scienze Fisiche, Università degli Studi di Napoli Federico II, I-80126 Napoli, Italy}
\affiliation{CNR-SPIN Napoli, Complesso Universitario di Monte Sant'Angelo, 80126 Napoli, Italy}
\author{R. Arpaia}%
\affiliation{CNR-SPIN Napoli, Complesso Universitario di Monte Sant'Angelo, 80126 Napoli, Italy}
\affiliation{Dipartimento di Scienze Fisiche, Università degli Studi di Napoli Federico II, I-80126 Napoli, Italy}
\affiliation{Quantum Device Physics Laboratory, Department of Microtechnology and Nanoscience, Chalmers University of Technology, SE-41296 Göteborg, Sweden}
\author{Y. Suzuki}
\affiliation{University of Tsukuba, Institute of Materials Science, Tsukuba, 305 Ibaraki, Japan}
\author{K. Kadowaki}%
\affiliation{University of Tsukuba, Institute of Materials Science, Tsukuba, 305 Ibaraki, Japan}
\author{T Bauch}
\affiliation{Quantum Device Physics Laboratory, Department of Microtechnology and Nanoscience, Chalmers University of Technology, SE-41296 Göteborg, Sweden}
\author{A. Tagliacozzo}
\affiliation{Dipartimento di Scienze Fisiche, Università degli Studi di Napoli Federico II, I-80126 Napoli, Italy}
\author{F. Tafuri}%
\affiliation{CNR-SPIN Napoli, Complesso Universitario di Monte Sant'Angelo, 80126 Napoli, Italy}
\affiliation{Dipartimento di Ingegneria Industriale e dell'Informazione, Seconda Universit\`a di Napoli, I-81031 Aversa (CE), Italy}
\author{F. Lombardi}%
\affiliation{Quantum Device Physics Laboratory, Department of Microtechnology and Nanoscience, Chalmers University of Technology, SE-41296 Göteborg, Sweden}
\date{\today}%

\begin{abstract}
In superconductor-topological insulator-superconductor hybrid junctions, the barrier edge states are expected to be protected against backscattering, to generate unconventional proximity effects, and, possibly, to signal the presence of Majorana fermions. The standards of proximity modes for these types of structures have to be settled for a neat identification of possible new entities. Through a systematic and complete set of measurements of the Josephson properties we find evidence of ballistic transport in coplanar Al-\BiSe-Al junctions that we attribute to a coherent transport through the topological edge state. The shunting effect of the bulk only influences the normal transport. This behavior, which can be considered to some extent universal, is fairly independent of the specific features of superconducting electrodes. A comparative study of Shubnikov - de Haas oscillations and Scanning Tunneling Spectroscopy gave an experimental signature compatible with a two dimensional electron transport channel with a Dirac dispersion relation.
A reduction of the size of the \BiSe\ flakes to the nanoscale is an unavoidable step to drive Josephson junctions in the proper regime to detect possible distinctive features of Majorana fermions.
\end{abstract}

\maketitle

\section{Introduction}

The understanding of how superconductivity propagates in metallic-like barriers has progressively become more and more comprehensive, taking advantage of the possibility of manufacturing a larger variety of interfaces and materials.  Examples of recent success come from a full description of superconductor-ferromagnet-superconductor (S-F-S) junctions\cite{Buzdin,GolubovKuprianov}, or from the possibility to include nano/mesoscopic features into a standard proximity effect\cite{CuevasdcSQUID,BergeretCuevas,Argaman}.
Novel flavors on the proximity effect are recently coming from the integration of nanowires, or quasi bi-dimensional systems, such as the edge states of topological insulators (TIs)\cite{Fu-Kane-Mele,vonOppen,Kovenhauven,Deng2012,Das2012,Half-Integer,Churchill2013,Finck2013}, as barriers. Here, well established and universally accepted concepts need to be revised. This is the case of the proximity effect through a TI barrier, where the transport through topologically protected edge states leads to a completely novel Josephson phenomenology, which should manifest neat fingerprints of Majorana fermions. These particles, which are their own antiparticles, are indeed expected to emerge in hybrid structures, involving the interface between a superconductor and a TI\cite{Fu-Kane-Mele,Nilsson2008,Tanaka2009}, or between a superconductor and a nanowire with strong spin orbit interaction \citep{Lutchyn2010,Oreg2010}. Superconducting hybrid structures are also considered  a crucial  step towards a topological quantum computer, which would be exceptionally well protected from errors, thanks to the non-Abelian statistics of Majorana fermions\cite{Nayak2008,Stern2010}.

Only a full understanding of proximity effect in superconductor - topological insulator structures can permit a clear identification of what comes from the particular morphology of the junction, and what is really unconventional in the sense of imprinting of topologically non trivial states. For example, the presence of topologically protected Majorana bound states (MBS) is expected to influence the current-phase relation (CPR) of a hybrid superconductor-topological insulator-superconductor (S-TI-S) Josephson junction\cite{Kitaev}. Phase sensitive measurements, such as the study of Shapiro steps and magnetic modulations in superconducting quantum interference devices (SQUIDs) and Josephson junctions, are good candidates to verify the presence of anomalous CPR in proximity devices with TI barriers \citep{Barone,Kang2000,Loder2008}.

Recent experiments have shown the first observation of a Josephson supercurrent through topological insulators \BiSe\cite{Morpurgo,StanfordTI,Zhang2011,Zareapour2012,Yang2012,Yang2012a} and Bi$_2$Te$_3$\cite{SdH1,Zareapour2012,Qu2012,Koren2012,VeldrostSQUID}. These Bi-based materials exhibit a residual bulk conductance that influences the electronic transport as a shunting channel, and that could``in principle'' mask signatures of induced superconductivity in the topological surface states. The contribution of the bulk is completely suppressed when strained HgTe is used as a barrier in S-TI-S hybrid junctions\cite{Oostinga2013}. However, also for these structures, the experimental study of the proximity Josephson effect has revealed only ordinary features characteristic of a conventional Josephson effect with a $\sin\varphi$ CPR. This seems to be consistent with the fact that the conditions to observe MBS related phenomenology are rather stringent, as recently pointed out in theoretical papers\cite{Snelder2013,Potter2013,VeldrostPRB}.

\begin{figure}[thbp]
	\centering
	\includegraphics[width=8.7cm]{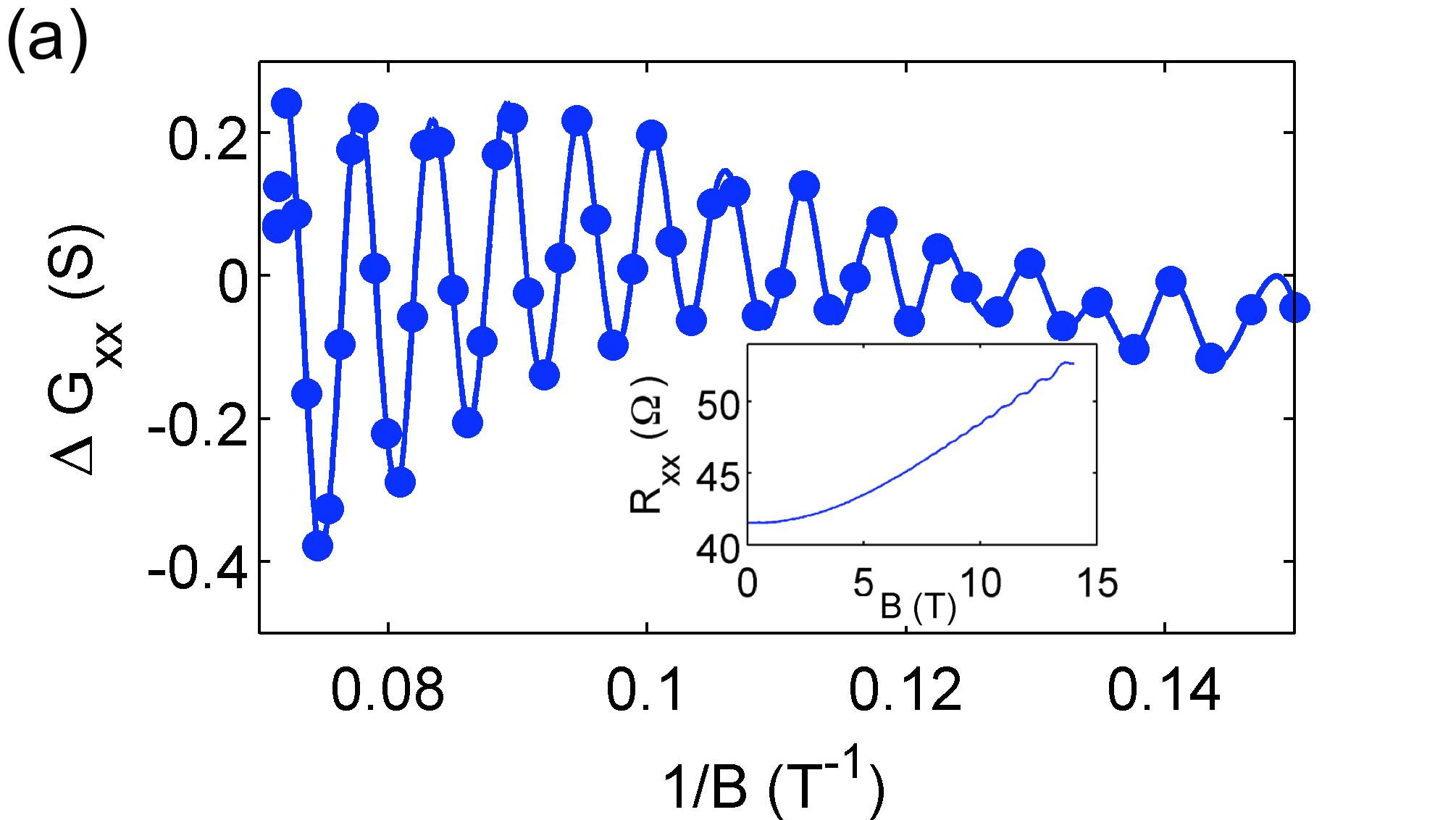}
	\includegraphics[width=8.7cm]{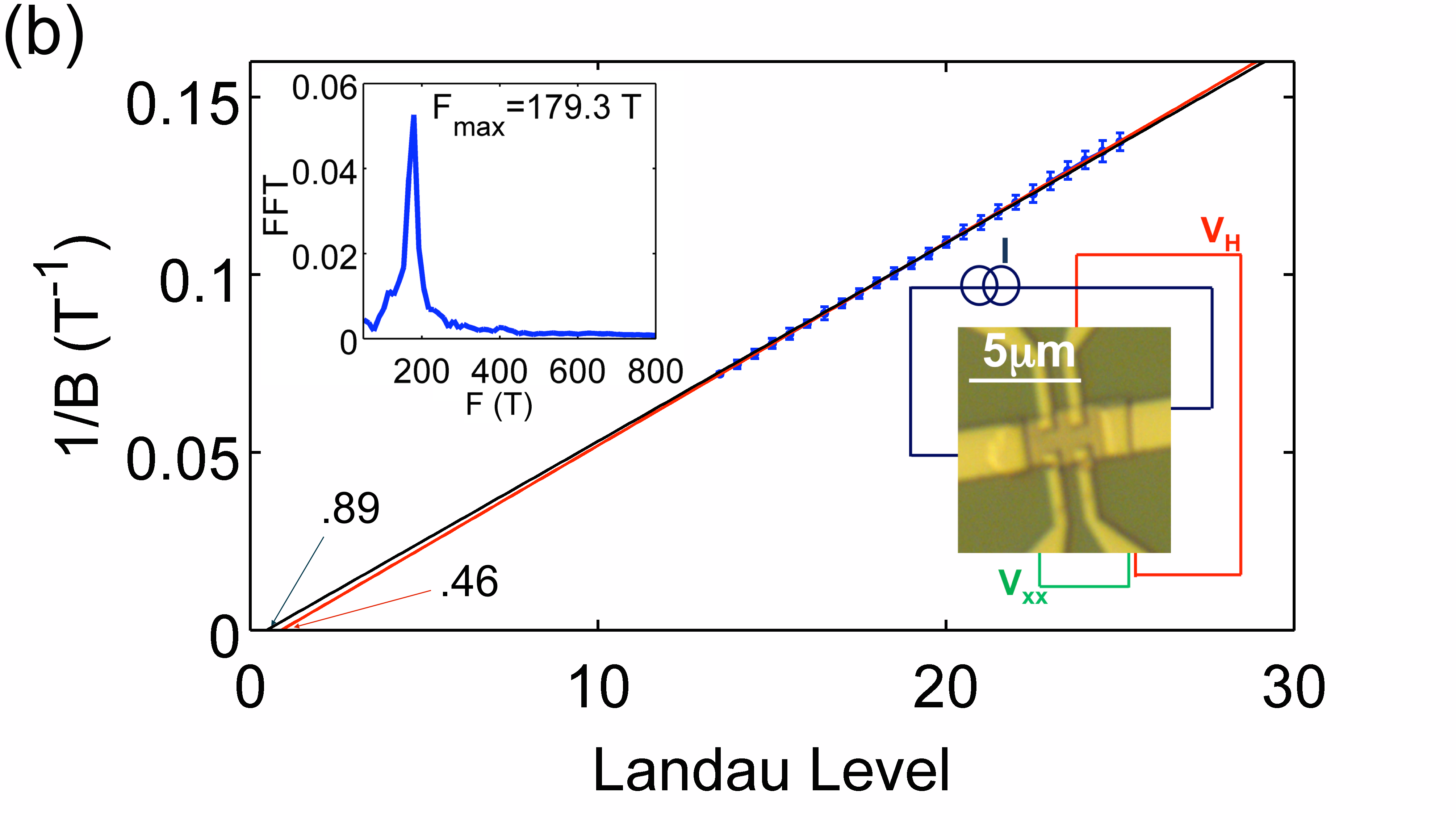}
	\caption{(Color online) (a) $G_{xx}$ after subtraction of the background as a function of the inverse of the magnetic field. The line is a spline interpolation of experimental data, and it is a guide for the eye. The inset shows $R_{xx}$ before the background subtraction. (b) Landau levels indexes inferred from conductance minima, and plotted as a function of $1/B$. The red line shows the fit obtained by fixing $F$ to the value F$_{max}$ extracted from the Fourier transform of $\Delta G_{xx}$ (see text for details), as shown in the top left inset. $F_{max}$~=~\unit{179.3}{\tesla} leads to an intercept $\beta =$~0.46. The black line is obtained while keeping both F and the intercept free for the linear fit: F~=~\unit{175.1}{\tesla} and the intercept $\beta$ is 0.89. The bottom right inset shows an optical picture of a flake contacted in a Hall bar configuration.}
	\label{Figure1}
\end{figure}

The present work, therefore, responds to the need of a better understanding of the properties of S-TI-S junctions and of their proximity regime. Our findings support the picture of a ballistic transport occurring  in edge states of \BiSe\ barriers independently of the superconducting electrodes, consistently with the conclusions of Ref.~[\onlinecite{SdH1}]. Our study is sustained by a complete independent characterization of the barrier material through transport measurements in Hall bar geometry and by scanning tunneling spectroscopy (STS) performed on the cleaved surface of the crystal. These complementary analyses provide a complete set of consistent parameters, which allow us to assess that topological protected surface states plays a crucial role in the Cooper pair transport in our systems.

\section{Magneto-conductance and Hall effect on Bi2Se3 flakes}

The devices presented in this paper are realized using single crystals of \BiSe\, fabricated by the melt-growth method\cite{Tanaka2012,Das2011}. Flakes of a thickness of $\simeq$ \unit{500}{\nano \meter} are exfoliated from the \BiSe\ crystal and reduced to \unit{30-90}{\nano \meter} through subsequent exfoliations\cite{Quintuple-Layer}.
The flakes are then transferred to a Si/SiO$_x$ substrate, and contacted with Al electrodes. To improve the adhesion with the Al electrodes a thin layer (\unit{3-5}{\nano \meter}) of Ti or Pt is deposited in situ on the surface of the flake, previously cleaned by an Ar$^+$ ion etching.

The observation of SdH (shown in Fig.~\ref{Figure1}(a)) reveals the presence of a high mobility two-dimensional (2D) transport channel with a carrier density of 4.1~-~4.8~cm$^{-2}$, in good agreement with data available in literature\cite{SdH1,Morpurgo}. The 2D nature of the transport has been confirmed through the measurement of the SdH as a function of the angle. This channel has been identified as the topological edge state of the crystal, observed in STM. 

\subsection{Shubnikov - de Haas oscillations}

We measured the Hall voltage $V_H$, across the device, as a function of the external magnetic field $B$ when a current $I$ is applied along the axis of the Hall bar. We have extracted the longitudinal $R_{xx}~=~V_{xx} / I$ and the transversal $R_{xy}~=~V_H / I$ resistance in a Hall bar geometry on exfoliated flakes, as shown in the inset of  Fig.~\ref{Figure1}(b). The values were recorded for a magnetic field sweeping from 0 to \unit{14}{\tesla} at \unit{2}{\kelvin} in a Quantum Design PPMS System. The corresponding values of $G_{xx}$ and $G_{xy}$ of the conductance tensor are evaluated with the equation set\cite{SdH2,sigma}
\begin{eqnarray}
G_{xx} &=&  R_{xx} / (R_{xy}^2 + R_{xx}^2)\nonumber\\
G_{xy} &=& -R_{xy} / (R_{xy}^2 + R_{xx}^2).
\label{tensor}
\end{eqnarray}

We performed measurements on six Hall bars and we observed Shubnikov-de Haas oscillations on five of them. The position in field of the Landau levels can be extracted from the minima of $G_{xx}$, signifying a complete filling of a number $n$ of Landau levels\cite{AndoRev}.  

In Fig.~\ref{Figure1}(b), we show a fan diagram where the Landau level indexes, determined from the minima of $G_{xx}$, are plotted as a function of the reverse their position in magnetic field $1/B_N$:
\begin{equation}
	(2N-1)=2 \left(\frac{F}{B_N} - \frac{1}{2} + \beta \right)
\end{equation}
where $N$ is the number of the minimum (corresponding to the number of the Landau level), $F$ is the frequency of the oscillations, given by the slope of the fan diagram. It is related to $n_{2D}$ by the Onsager's relation\cite{AndoRev}
\begin{equation}
	F = \frac{1}{2\pi}\left( \frac{\hbar c}{2 \pi e} \right) \pi k_F^2 = \left( \frac{\hbar c}{e} \right) n_{2D}.
	\label{Onsager}
\end{equation}
The linear fitting of the data allows us to extract the two-dimensional carrier density of the surface $n_{2D}$ and to infer the existence of a zero energy Landau level, proper of a Dirac material.
The value of $n_{2D}$ falls in the \unit{4.1 - 4.8 \cdot 10^{12}}{cm^{-2}} range, which confirms the high reproducibility of 2D edge state.
The value of the intercept $\beta$ of the fan diagram presented in Fig.~\ref{Figure1}(b), instead, gives information about the presence of a zero energy Landau level, expected in the case of Dirac electrons\cite{grapheneSdH,ZhangGraphene,grapheneSdH3}. For Dirac electrons, the Berry phase, $\varphi_B~=~2\pi \beta$ is equal to $\pi$ so the expected value of the intercept $\beta$ is 1/2. However, the conclusions of this analysis can strongly depend on the procedure\cite{AndoRev} used to extract the intercept from the fan diagram. To obtain the linear fit of the data in Fig.~\ref{Figure1}(b), we have fixed the slope of the fan diagram to the value of the frequency F$_{max}$~=~179.3~T extracted from the peak position of the Fourier transform analysis of the $G_{xx}$. We have then obtained an intercept  equal to 0.46, which leads to $\varphi_B\simeq \pi$, compatible with the presence of Dirac electrons. Alternatively, by keeping as free parameters both F and the intercept, we get a value close of 0.89 and a relative phase $\varphi_B\simeq 0$ which is instead in agreement with a conventional 2D electron gas. However, both values obtained for the intercept can be accounted for within the error on F extracted from the Fourier transform of the $G_{xx}$ data (see inset Fig.~\ref{Figure1}(b)).
This uncertainty is quite typical when the value of the magnetic field is not high enough to populate low order Landau levels (close to n~=~0). In most cases, the population of low order  Landau levels is reached for field of the order of 30~T\cite{SdH1}. As a consequence, the value of the phase extracted at lower fields (in our case 14~T), can be misleading, as discussed in literature\cite{SdH3D,OngPRB,SdH1,AndoRev}. In our sample, we have also performed STM measurements, a technique which allows to detect in a reliable way the presence of Dirac electrons (see section below).
The study of SdH as a function of the angle, presented in the next section, confirms the 2D nature of the SdH, ruling out the possibility that the oscillations arise from a bulk three-dimensional (3D) transport channel\cite{SdHtrivial}. Hall measurements performed at low fields give evidence of the presence of two transport channels, active in the bulk and at the edge of the crystal, respectively.

\subsection{SdH measurements as a function of the angle}

Shubinkov - de Haas oscillations are not an exclusive feature of 2D systems, but they can also be detected in 3D systems under some conditions, due to the formation of Landau bands\cite{SdHtrivial,SdH3D}.
\begin{figure}[htbp]
	\centering
	\includegraphics[width=8.5cm]{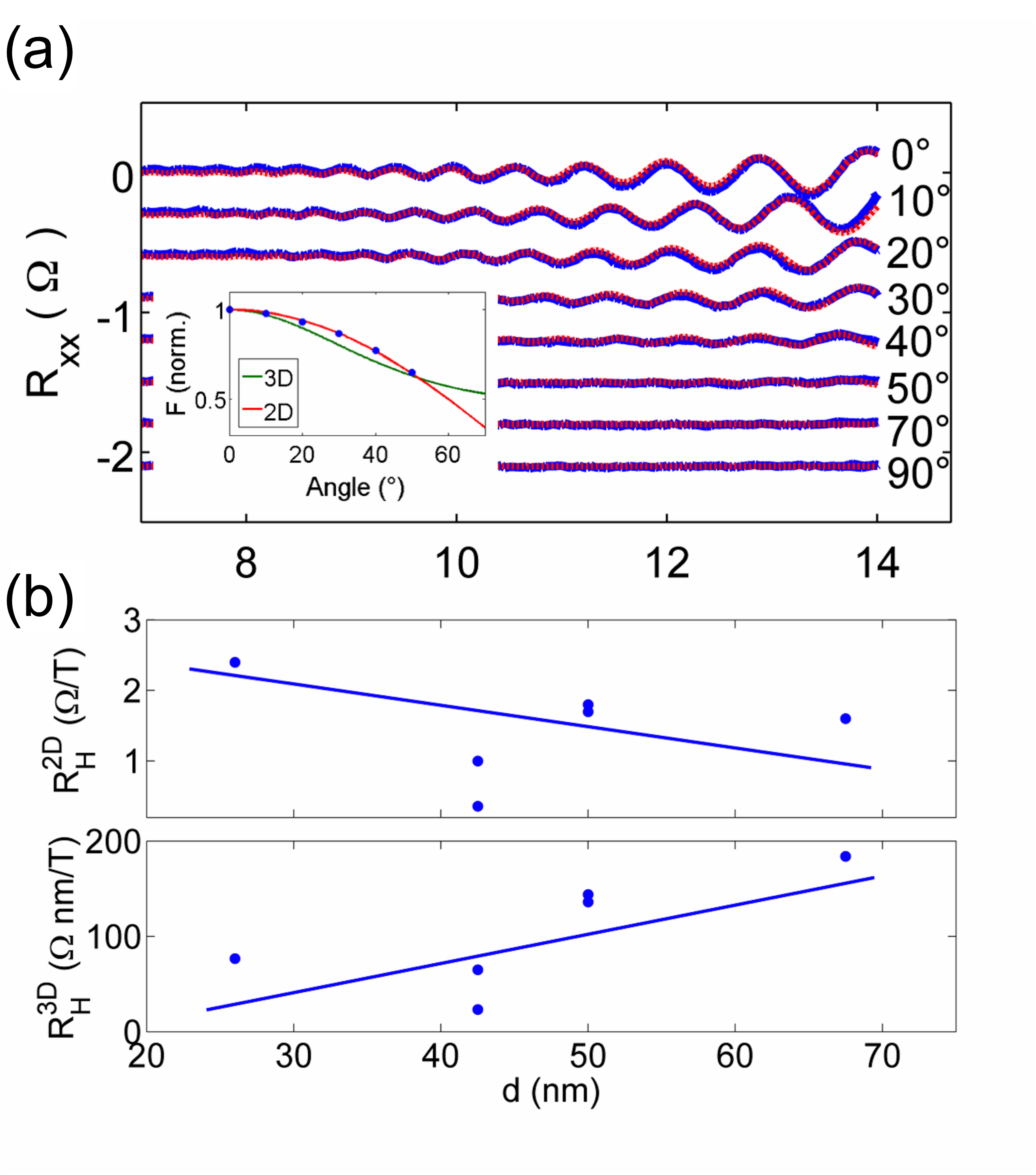}
	\caption{(Color online) (a) Shubnikov-de Haas oscillations as a function of the magnetic field ($B$) at different angles $\theta$ between the direction of $B$ and the plane of the current transport. The magnetoresistance background has been subtracted. Curves for various $\theta$ have been shifted for clarity. Experimental data are in blue, while the red lines are the fits using Eq.~(\ref{SdHmodel}). The inset shows the values of the oscillation frequency inferred from the fit at different angles $\theta$ (blue points). Here the red line is the co-sinusoidal behavior while the dashed green line is the expected dependence in the case of an ellipsoid shaped Fermi surface (two axes of the ellipsoid are equal and the third axis is 2 times bigger). This is the expected shape of the Fermi surface for the bulk band structure of \BiSe\cite{Stordeur1992,Mishra1997} (b) 2D and 3D Hall coefficients extracted from the linear fit of the $R_{xy} (B)$ curve between \unit{-5}{\tesla} and \unit{5}{\tesla}. The two coefficients show a clear trend with the flake thickness. A dependence of both 2D and 3D coefficient indicates a parallel transport through the bulk of the crystal and the 2D edge state.}
	\label{Figure2}
\end{figure}
A support to the 2D nature of SdH oscillations can be provided by a study of the behavior of $F$ as a function of the angle $\theta$ between the magnetic field and the surface of the sample. Specifically in the case of a 2D transport, $F$ changes as $1/\cos\theta$, while a 3D system shows deviations from this dependence, especially at high angles as shown in the inset of Fig.~\ref{Figure2}(a). 

We have measured $R_{xx}$ as a function of the magnetic field for different angles $\theta$. A set of measurements is shown in Fig.~\ref{Figure2}(a). The oscillations have been fitted by the Lifshiz-Kosevich model\cite{AndoRev}
\begin{equation}
	R_{xx} \propto \frac{\lambda}{\sinh\left(\lambda\right)}\exp\left({-\frac{\pi}{\omega_c \tau_c}}\right) \cos\left[{2\pi\left({\frac{F}{B} + \frac{1}{2} + \varphi'_B}\right)}\right]
\label{SdHmodel}
\end{equation}
where $\lambda~=~2 \pi^2 k_B T / \hbar \omega_c$, $\omega_c~=~eB / cm_c$ is the cyclotron frequency, $\tau_c$ is the scattering time and $m_c$ is the cyclotron mass. $F$ is defined by the Onsager's relation (Eq.~\ref{Onsager}) and $\varphi'_B$ is a phase shift related to the presence of a Landau level at zero energy.

In Fig.~\ref{Figure2}(a) we show the modulation of the frequency of the resistance oscillations as a function of the angle \footnote{Shubnikov-de Haas oscillations have been studied on various Bi compounds with various interpretations\cite{SdH1,SdH2,SdHbis,SdH3,SdH4,SdH6,SdH8}. Quite recently SdH detected on \BiSe\ flakes were attributed to an effect of a possible layered bulk material, having therefore a cylindrical Fermi surface (implying a cosine dependence of the frequency of the oscillations as a function of the angle)\cite{BiSe-bulk}. In this work the authors also extracted a $\varphi_B$~=~0 phase, in agreement with the absence of Dirac electrons.}. The modulation of $F$ as a function of the angle $\theta$ follows a cosine curve up to at least 50°, as expected in the case of a 2D transport\cite{SdH2}, and differently from the ellipsoid shaped Fermi surface expected for the bulk band structure of the \BiSe\cite{SdHtrivial,Barkeshli}.

\subsection{Linear Hall Regime}

While high field measurements are able to reveal the presence of a high mobility 2D channel, low field measurements are dominated by the shunting effect of the bulk of the crystal. In the linear regime, the Hall coefficient is independent of the sample geometry\cite{SdHbis}, and it is simply related to the carrier density of the device. In the case of a 2D system the Hall coefficient is defined as $R_H^{2D}~=~V_H~/~B I$, while in 3D systems this quantity should be normalized to the thickness $d$ of the Hall bar ($R_H^{3D}~=~d V_H~/~B I$)\cite{SdHbis}. Clearly, $R_H^{2D}$ and $R_H^{3D}$ cannot be simultaneously independent of the sample thickness. One can infer if the transport takes place in the bulk of the flake or in the 2D edge states by studying the thickness dependence of the 2D and 3D Hall coefficients.
In our case, in spite of the pronounced data scattering a dependence of both the 2D and the 3D Hall coefficients on the thickness $d$ can be observed, as shown in Fig.~\ref{Figure2}(b) demonstrating the presence of two channels contributing simultaneously to the transport, namely the bulk and the surface\cite{SdHbis,Kim2011}. 

\section{STM measurements of Bi$_2$Se$_3$}

\begin{figure}[bhtp]
	\centering
	\includegraphics[width=8cm]{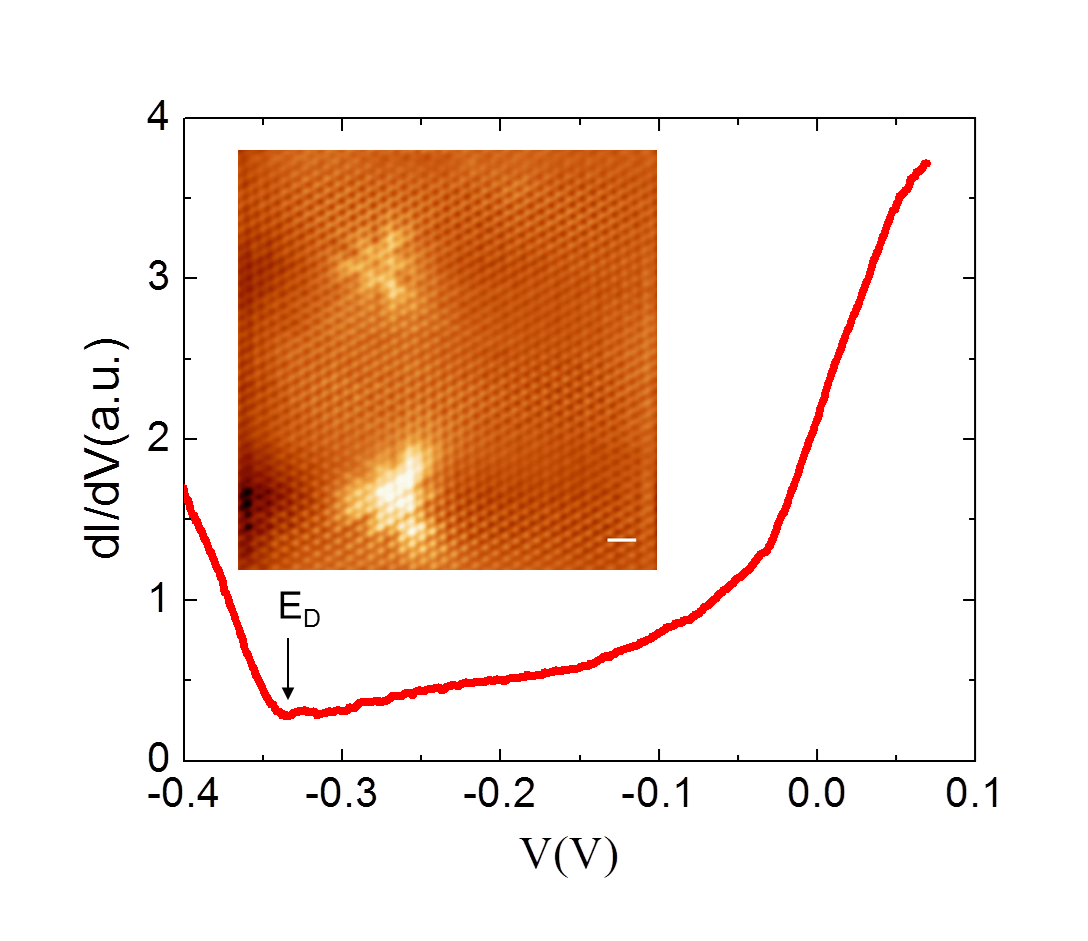}
	\caption{Tunneling conductance spectrum acquired at the surface of a \BiSe\ crystal. The spectrum has been measured with a bias modulation amplitude $V_{mod}$~=~\unit{2.5}{\milli V_{rms}}. The V-shaped spectrum is consistent with the Dirac dispersion with the Dirac point at $V$~=~\unit{-350}{\milli \volt} (minimum of the spectrum indicated by the arrow and labeled as $E_D$). The inset is an atomic resolved topographic image showing triangular shaped defects. The imaging conditions are: $V$~=~+0.1~V, $I$~=~10~pA, scan area is 14.5~nm~x~14.5~nm (scale bar is 1~nm).}
	\label{STM1}
\end{figure}

\begin{figure}[bhtp]
	\centering
	\includegraphics[width=8cm]{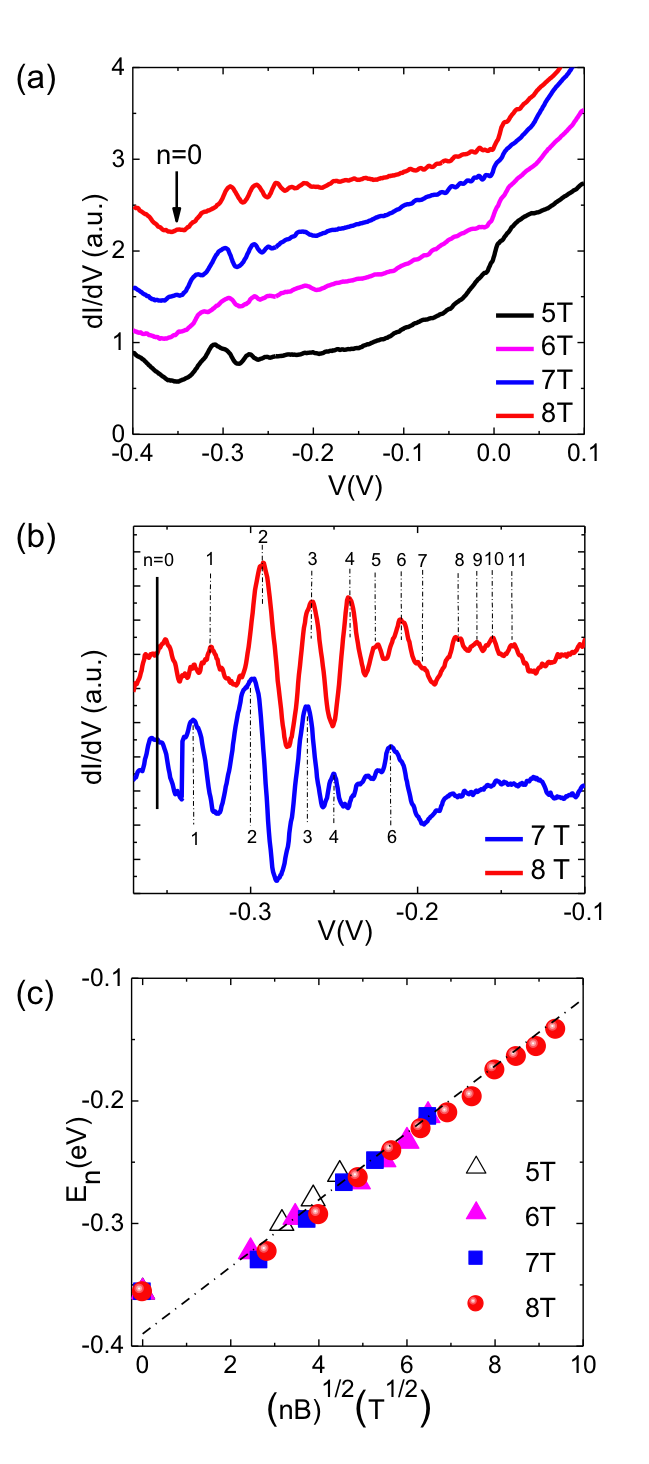}
	\caption{(a) Tunneling conductance spectra acquired in the presence of a magnetic field applied perpendicular to the sample surface. The spectra at different fields have been shifted for clarity. The tunneling spectra above 5~T show a series of peaks associated with the formation of Landau levels. (b) dI/dV spectra at 7~T (in blue) and 8~T (in red) of Fig.~4(a). After the subtraction the Landau levels are clearly visible and they can be labeled. (c) Landau levels energy $E_n$ as a function of $\sqrt{nB}$. The dashed line is the linear fit of the data.}
	\label{STM2}
\end{figure}

The existence of a Dirac cone at the surface of our crystals has been verified by Scanning Tunneling Spectroscopy (STS) measurements (Fig.~\ref{STM1} and Fig.~\ref{STM2}). \BiSe\ was studied with an ultra high vacuum (UHV) STM (Unisoku USM-1300).
Single crystals were cleaved at room temperature in UHV and immediately transferred to the STM scanner at low temperature. Differential conductance spectra ($dI/dV(V)$), which reflect the local density of states, were obtained using a standard lock-in technique at a temperature of 2~K. 

In Fig.~\ref{STM1} a typical conductance spectrum is reported. The spectrum exhibits a V-shape with a minimum at $V$~=~\unit{-350}{\milli \volt}, representing the Dirac point. The presence of the Dirac point located below the Fermi level has already been reported in literature\cite{Hanaguri2010,Cheng2010,Hansan-Kane,Hsieh} and it has been attributed to the presence of defects, predominantly Se vacancies. STM topographic images (inset of Fig.~\ref{STM1}) reveal the presence of triangular-shaped defects with a spacing of few nanometers that may represent either Se vacancies and/or defects in the Bi plane.

STS spectra, acquired in a magnetic field applied perpendicular to the sample surface, present not equally spaced peaks that are clearly visible above \unit{5}{T} (Fig.~\ref{STM2}(a)) and that are in agreement with Landau energy levels. Since the Landau levels are a small perturbation on a large background the Landau levels are presented in Fig.~\ref{STM2}(b) with the background subtracted, for applied magnetic fields of 7 and \unit{8}{T}. It can be shown that, in Dirac systems, the energy dependence of the Landau levels, follows the aperiodic dispersion relation given by\cite{RevModPhys.81.109}: $E_n~=~E_D~+~\text{sgn}(n)v_F \sqrt{2e\hbar \lvert n\rvert B}$, where $E_n$ is the energy of the $n$th Landau level, $n~=~(0, \pm 1, \pm 2... \pm\!n)$ is the Landau level index, $E_D$ is the Dirac point, $v_F$ the Fermi velocity, $\hbar$ is the Plank constant and $B$ the magnetic field\cite{Zheng2002}. A plot of $E_n$ versus $\sqrt{nB}$ reveals a reasonable good linear behavior above the Dirac point as expected for the dispersion of a topological insulator (Fig.~\ref{STM2}(c)). Slight deviations from the linearity can be attributed to the tip-gating effect\cite{Cheng2010}.
No Landau levels were observed below the Dirac point. This is consistent with previous reports on \BiSe\ crystals\cite{Cheng2010} and it can be attributed to a coupling to the bulk valence band located just below the Dirac point. An important feature is the presence of a peak $n~=~0$ at the Dirac point that is independent of the applied magnetic field.   
For a conventional two-dimensional electron gas system with a parabolic dispersion such a zeroth Landau level is not permitted.

\section{Proximity effect in Josephson junctions and d.c. SQUIDs}\label{Proximity}

Josephson junctions and Superconducting Quantum Interference Devices (SQUIDs) have been fabricated, using the crystals of \BiSe\ described above.

Recently, superconductivity has been induced in \BiSe\ flakes both by Nb and Al electrodes. In Nb/Bi$_2$Te$_3$/Nb junctions evidence of a ballistic regime, with a conventional CPR, has been given\cite{SdH1,VeldrostPRB,VeldrostSQUID}. This has been interpreted as sign of a prevalent transport of non-trivial two-dimensional surface state. Previous experiments on Al/\BiSe/Al Josephson junctions\cite{Morpurgo,StanfordTI,StanfordNano}  have given evidence of Josephson coupling. In the paper by \citet{StanfordTI}, critical current dependence on the magnetic field has been shown to have unconventional behavior, later explained to be related to the presence of Pearl vortices in a long junction regime\cite{StanfordNano}. \citet{Morpurgo} demonstrated gate tuning of the normal and superconducting properties of the junction. These results are summarized in Table \ref{table}

\begin{table*}[htbp]
\begin{tabular}{| p{1.7cm} | c | c | c | c | c | c | c | p{2.4cm} |}
	\hline
	Reference & Junction &  $L$ (nm) & $I_c$ ($\mu$A) & $e I_c R_n / \Delta$ & $\gamma$  & $\mu$ (m$^2$/Vs) & $n$ (\unit{10^{12}}{cm^{-2}}) & Notes \\
	\hline
	\hline
	This work (Ti buffer layer) & Al/\BiSe & 400 & 0.228 & 0.011 & 0.6 & - & 4.1 - 4.8 & Fig \ref{Figure7}\\
	\hline
	This work (Pt buffer layer) & Al/\BiSe & 300 & 1.67 & 0.086 & 0.9 & - & 4.1 - 4.8 & Fig \ref{Figure5},\ref{Figure4},\ref{Figure7}b (inset)\\
	\hline
	\citet{SdH1} &Nb/Bi$_2$Te$_3$ & 50 & 18 & 0.02 & - & 0.8 &1.2 & Ballistic proximity effect \\
	\hline
	\citet{Morpurgo} & Al/\BiSe & 400 & 0.3 & 0.073 - 0.093 & - & 0.5 & 1 - 5 & Gate tunability of $n$ and $I_c$ \\
	\hline
	\citet{StanfordTI}  & Al/\BiSe & 45 & 0.850 & 0.067 - 0.267 & - & - & - & Anomalous Fraunhofer patterns \\
	\hline
	\citet{Half-Integer} & Nb/InSb & 30 & 0.45 & 0.13 & $>$0 & - & - & Anomalous Shapiro steps \\
	\hline
	\citet{Oostinga2013} & Nb/HgTe & 200 & 3.8 & 0.19 & 0.5 & 2.6 & 0.5 & No bulk losses \\
	\hline
\end{tabular}
\caption{Comparison of the parameters of two of the junctions realized in this work with those available in the literature.}
\label{table}
\end{table*}

The body of our results (presented below) confirms, for Al/\BiSe/Al junctions, the dominant role of two dimensional surface states to carry supercurrent, which can be considered to some extent universal. \BiSe\ crystals, which are produced in quite different conditions and whose surfaces are treated in quite different manners\cite{VortexNOI}, behave as ballistic barriers independently of the exact interface with the superconducting electrodes.

Measurements have been taken at low temperature in a pseudo-four-point configuration, by anchoring the sample to the cold stage of a $^3$He/$^4$He Oxford ``Kelvinox" dilution refrigerator. As for a complementary analysis a $^3$He evaporation refrigerator Oxford ``Heliox" has also been used. Both cryogenic systems were equipped with resistor-capacitor low pass filters, with a cut-off frequency of 1.6 MHz, and Cu powder filters at low temperature. A room temperature stage of low pass filters has also been used. All the electronics have been designed to minimize the thermal noise\cite{LongobardiMicrostrip}. Further details on the electronics are described elsewhere\cite{Longobardi}.
The cold stage of the cryostat was equipped with a \unit{10}{\milli \tesla} superconducting coil (\unit{300}{\milli \tesla} in the `Heliox'), to apply an external magnetic field perpendicularly to the junction plane (see the sketch in the inset of Fig.~\ref{Figure4}(a)). The sample was also placed at about \unit{1}{cm} below a microwave antenna, therefore an rf signal with tunable frequency and amplitude could be shined on the sample.

The magnetic pattern demonstrates classical Fraunhofer features, as confirmed by our measurements and, as a further check, by the magnetic modulations of a Al/\BiSe/Al SQUID and a reference Al SQUID. No peculiar unconventional effect has been observed (see Fig.~\ref{Figure4}).  The same conclusions of no unconventional behavior induced by the \BiSe\ barrier apply to properties of the junctions in presence of rf radiation\cite{Half-Integer} (see Fig~\ref{Figure5}).

The temperature dependence of the critical current $I_c$ is the most revealing measurement of the ballistic regime of the junction, as discussed in Sec.~\ref{sezioneT}  (see Fig.~\ref{Figure7}).

The devices had a typical separation between electrodes of \unit{300 - 400}{\nano \meter},  they showed a metallic-like resistivity down to \unit{50}{\kelvin}, and an almost temperature independent resistance below this temperature. At \unit{1.1}{\kelvin} a transition to a superconducting state is observed. The proximity effect is induced in \BiSe\ through the Al superconducting leads. Our devices showed a critical current ranging between \unit{50}{\nano \ampere} and \unit{1.5}{\micro \ampere} at \unit{300}{\milli \kelvin}. The value of the critical current is mainly limited by the transparency of the interfaces. Despite some spread in the junction parameters, due to the quality, surface treatment and aging of the samples, the same conclusions discussed below are of general validity. High transparency interfaces were achieved using a Pt interlayer between the Al electrode and the \BiSe\ flake. Devices with lower barrier transparency have been achieved using a Ti interface. The Current-Voltage characteristics (IV curves) of our devices show a non hysteretic behavior, well described by the Resistively Shunted Junction (RSJ) model\cite{Barone}. 

\subsection{Microwave Field in Josephson Junctions}

To study the nature of dissipation-less transport in our junctions, we performed measurements of IV curves in the presence of microwave irradiation. We have detected Shapiro steps in the interval of \unit{1.5 - 4}{\giga \hertz}, confirming the presence of the a.c. Josephson effect. 
\begin{figure}[htbp]
	\centering
	\includegraphics[width=8.5cm]{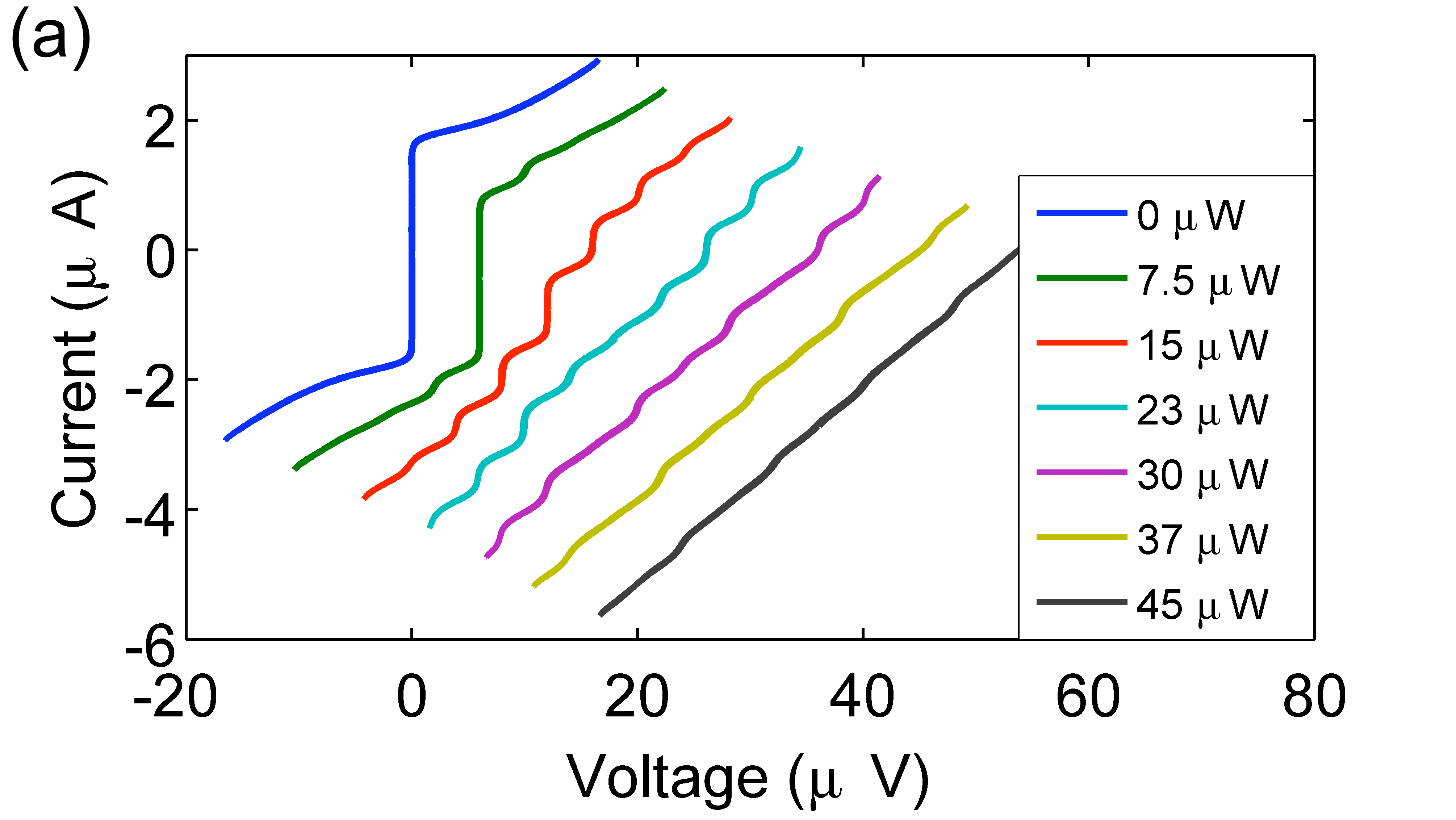}\\
	\includegraphics[width=8.5cm]{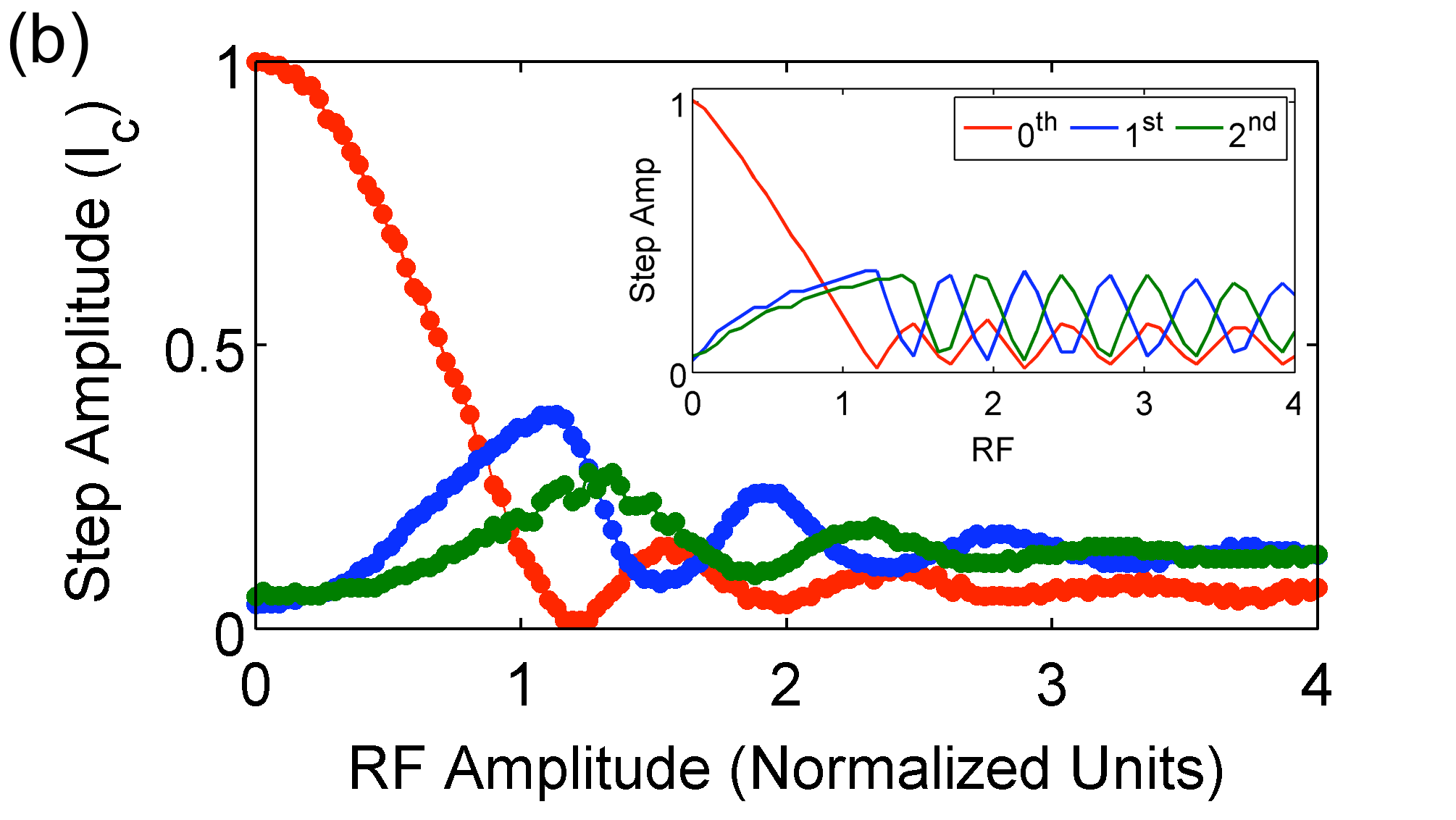}\\
	\caption{(a) IV curves of a Josephson junction at the temperature of \unit{300}{\milli \kelvin} for different power of the applied microwaves at the output of the microwave source, with a frequency of \unit{2}{GHz}. The curves have been shifted for clarity. (b) Experimental and simulated (inset) data showing the modulation of the critical current and of the amplitude of the 2 first Shapiro steps as a function of the applied microwave power. Here we have assumed a conventional $I=I_c\sin\varphi$ CPR.}
	\label{Figure5}
\end{figure}
In Fig.~\ref{Figure5}(a), typical IV curves with an increasing microwave radiation power are displayed. Since our junctions are current biased, and their characteristic voltage $I_c R_N$ is much bigger than the microwave frequency $\hbar \omega_{RF}/2e$ (nominally the dimensionless parameter $\eta = \hbar \omega_{RF} / 2 e I_c R_N$ is 0.22), we are in a regime where the modulation of the Shapiro steps does not follow the Bessel-like dependance on microwave amplitude, relevant for voltage biased JJs\cite{Barone}. The modulation of the lowest order steps was then numerically simulated using equation\cite{Shapiro_num}
\begin{equation}
	\alpha_0 + \alpha_1 \sin\left(\eta \tau\right) = \frac{d \varphi}{d \tau} + \sin\varphi
	\label{ShapiroSim}
\end{equation}
where $\alpha_0 = I/I_c$, $\alpha_1 = I_{RF}/I_c$ are the d.c. and a.c. component of the current bias, normalized by the critical current and $\tau = \Omega t$ the normalized time with $\Omega = 2eI_{c}R_N/\hbar$.
Shapiro steps modulations measured on \BiSe\ junctions have been efficiently reproduced using a $\sin \varphi$ current phase relation. In Fig.~\ref{Figure5}(b), the measured modulation of the first 3 Shapiro steps are compared with numerical simulation, obtained solving Eq \ref{ShapiroSim} (inset of Fig.~\ref{Figure5}(b)). The good agreement between data and simulations strongly supports a Josephson transport with a conventional CPR.

\subsection{Magnetic Field Features in Josephson Junctions and SQUIDs}

\begin{figure}[htbp]
	\centering
	\includegraphics[width=8.5cm]{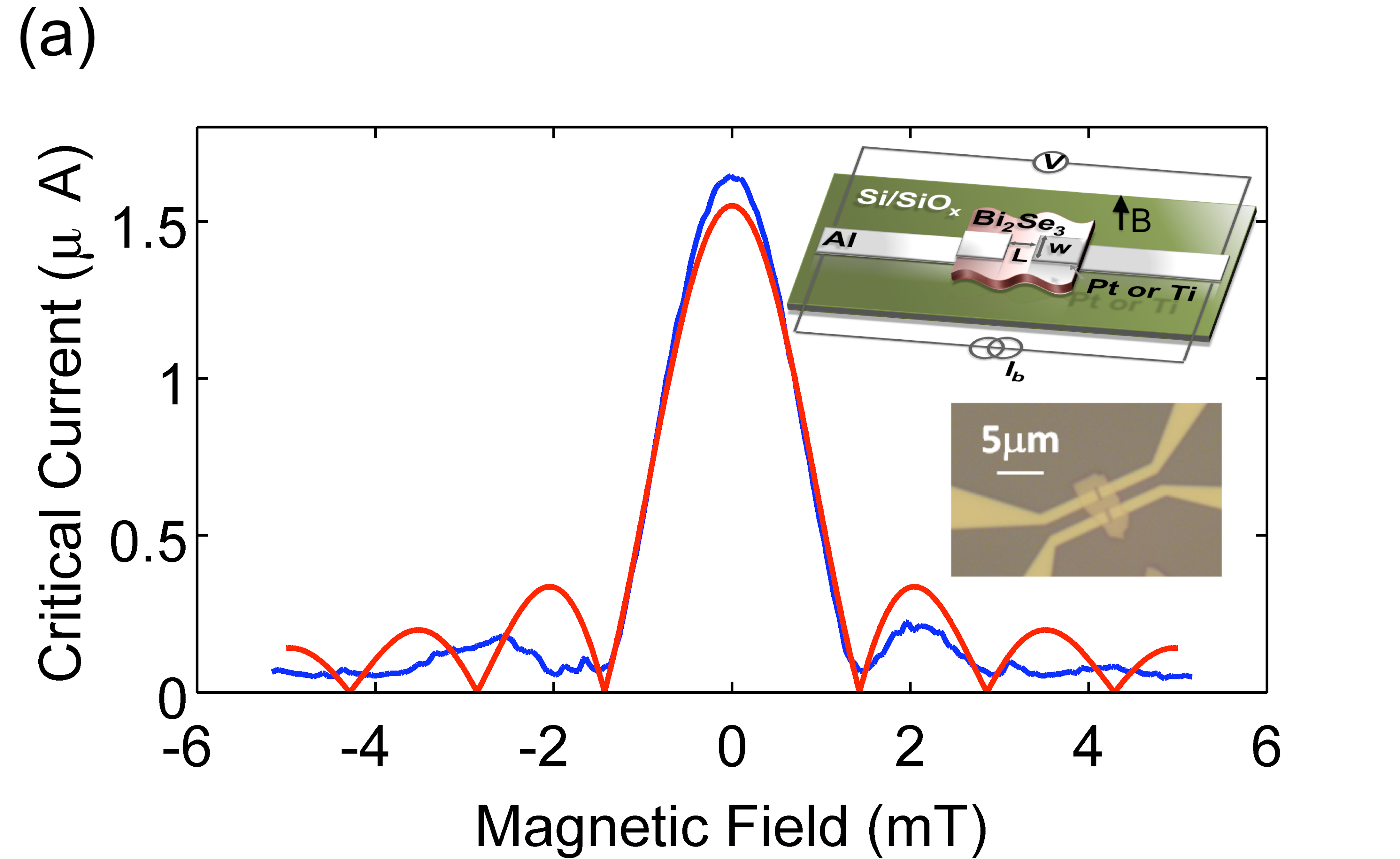}\\
	\includegraphics[width=8.5cm]{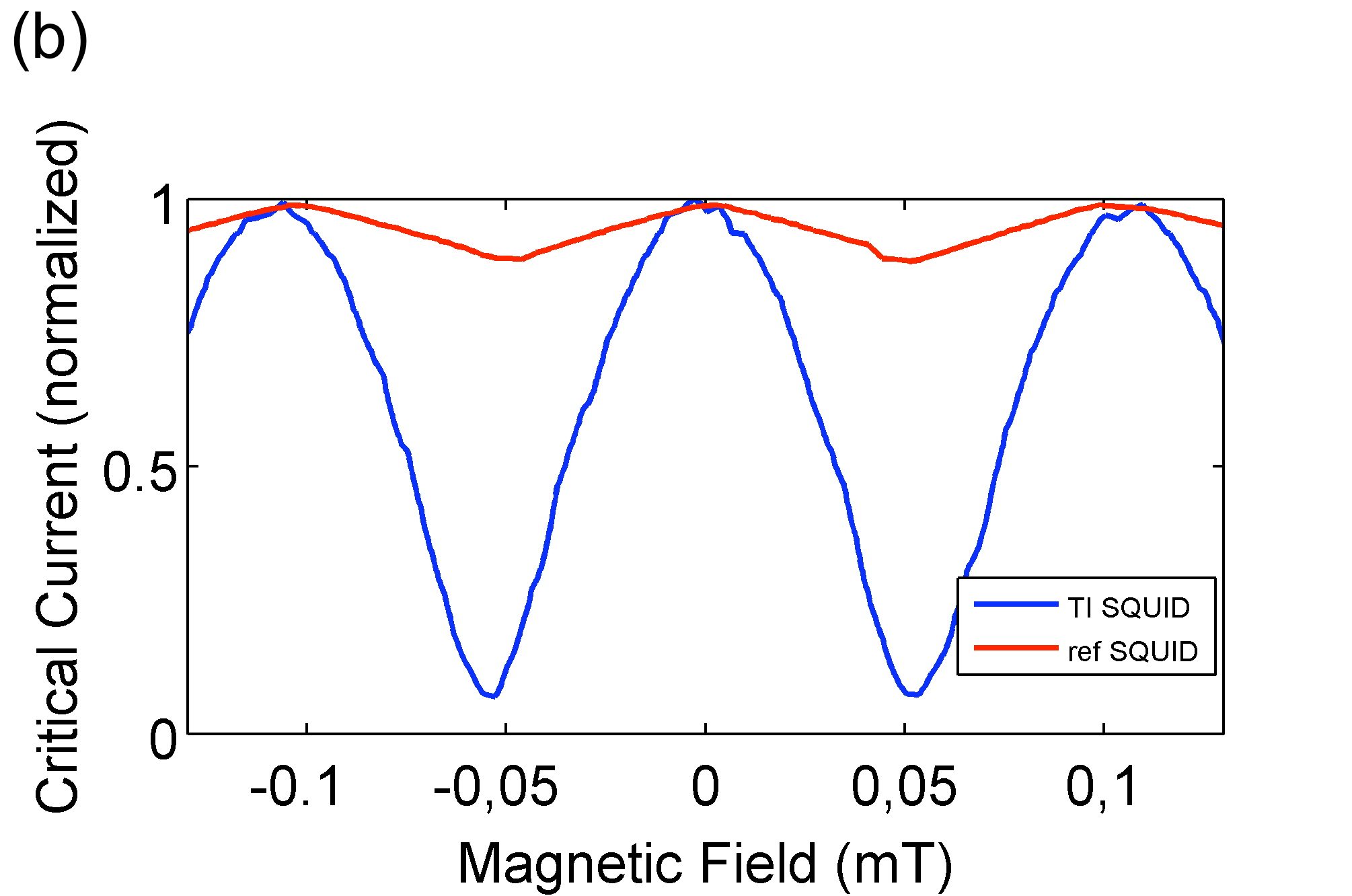}\\
	\caption{(a) Critical current as a function of the external magnetic field in a Al/\BiSe /Al Josephson junction at \unit{300}{\milli \kelvin}. The red line is the reference curve appropriate for a small junction with a uniform critical current density. The inset shows a sketch of the device and an optical image.
	(b) Critical current modulations as a function of the magnetic field of a  Al/\BiSe /Al SQUID (blue curve), compared to that of a reference SQUID (red curve). Both measurements have been performed at \unit{300}{\milli \kelvin}. The two devices have modulations with the same periodicity. The different modulation depth is related to the different value of the kinetic inductance of the loops.
	}
	\label{Figure4}
\end{figure}

Possible anomalies in the CPR can be also detected by studying the modulations of the critical current of junctions and SQUIDs as a function of an external magnetic field, via the d.c. Josephson effect \citep{Barone,Kang2000,Loder2008,VeldrostPRB}. We have measured conventional Fraunhofer patterns in Josephson junctions with \BiSe\ barrier, see Fig.~\ref{Figure4}(a). 
Indeed the measured period of the magnetic field pattern is in reasonable agreement with the expected period, considering a conventional CPR\cite{Rosenthal1991}. Also, the London penetration length in the thin aluminum electrodes could be higher than their bulk value\cite{Cohen1968,Behroozi1974}, which can make the real area of the junction larger than expected. We also measured the modulation of the critical current as a function of the magnetic field for d.c. SQUID with the same \BiSe\  barrier, as presented in Fig.~\ref{Figure4}(b). For comparison we have measured the period of the SQUID oscillations of a bare Al reference SQUID, realized on the same chip (also shown in Fig.~\ref{Figure4}(b)).
The Al/\BiSe\ SQUID and the reference device show the same modulation period, which further supports the existence of a conventional CPR.

\subsection{Temperature Dependence of the Critical Current}
\label{sezioneT}

\begin{figure}[htbp]
	\centering
	\includegraphics[width=8.5cm]{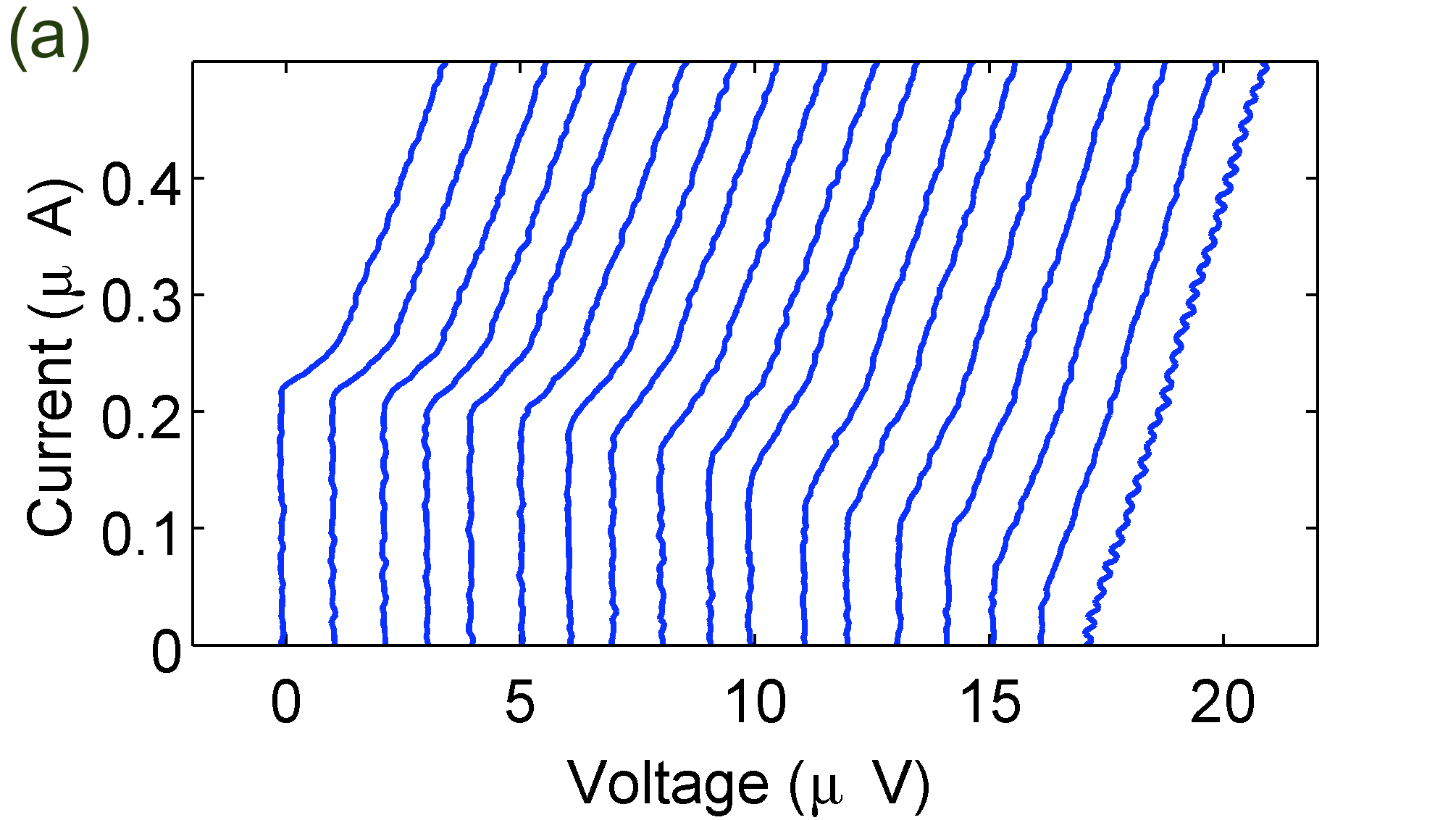}\\
	\includegraphics[width=8.5cm]{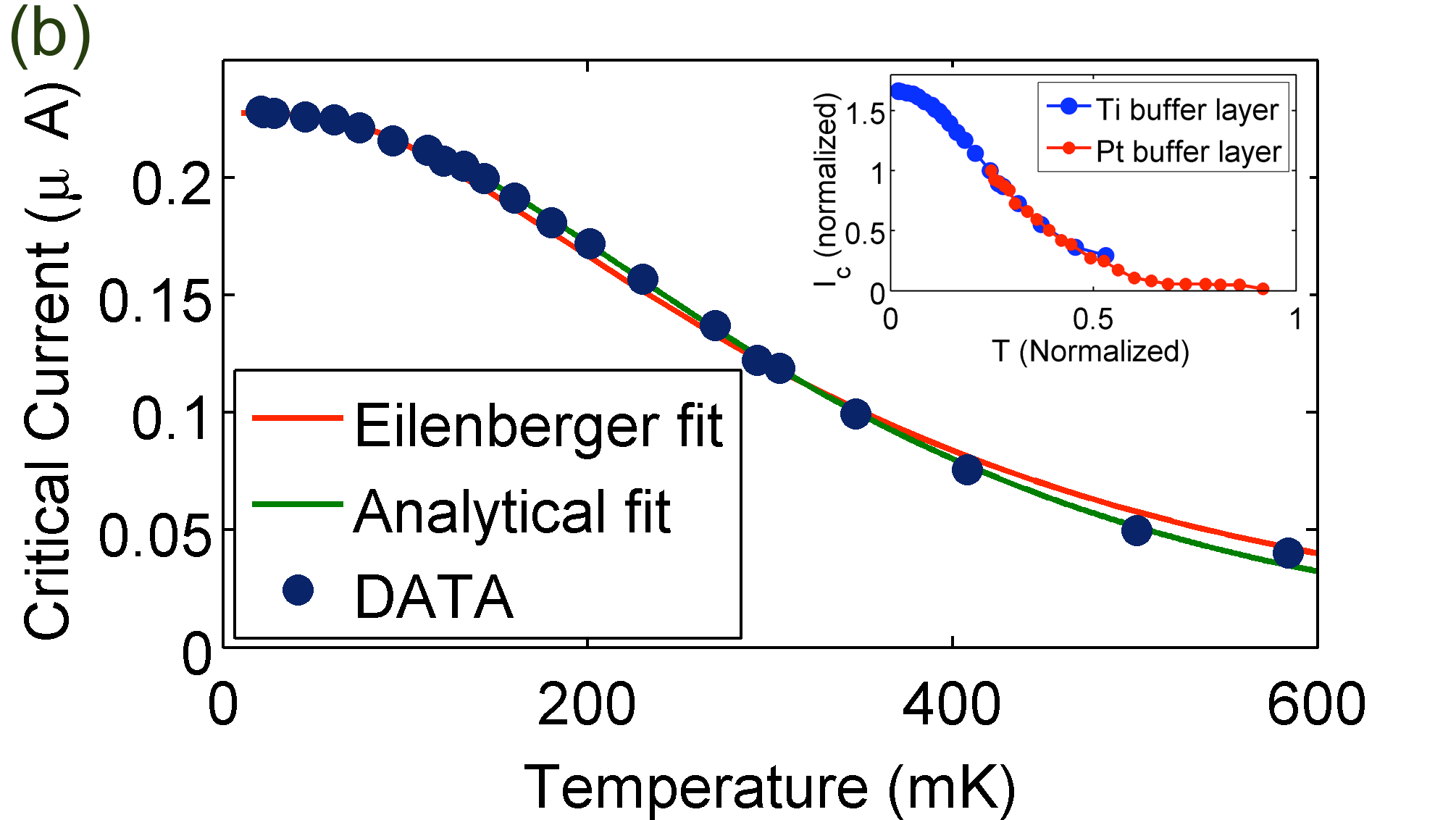}\\
	\caption{(a) IV curves of a Josephson junction as a function of the temperature. (b) Critical current of a Josephson junction as a function of the temperature. The analytical fit of the data to Eq. \ref{ModelProc} leads to a Thouless energy of about $E_{th}$~=~\unit{100}{\micro e \volt} (shown in green). The numerical Eilenberger fit is shown in red. In the inset the critical current as a function of temperature is reported, in normalized units, for a junction with a Ti buffer layer and a junction of Pt buffer layer. Both devices show the same trend.
	}
	\label{Figure7}
\end{figure}

In Fig.~\ref{Figure7}(b), we report the critical current of one of the devices as a function of the temperature, down to \unit{20}{\milli \kelvin}. The device was fabricated using a Ti interface, and shows a critical current of about \unit{230}{\nano \ampere}. In the inset we have also rescaled the critical currents of two samples with different buffer layers (Pt and Ti), showing a remarkable collapse onto the same curve.

We have performed a fit procedure of the $I_c$ vs $T$ curves  using the equation\cite{Kresin1986,Lehnert1999}
\begin{equation}
	I_c(T) \propto \sqrt{T}e^{-\frac{2 \pi k_B T}{E_{th}}}
	\label{ModelProc}
\end{equation}
 where $E_{th}$ is the Thouless energy. A value of the Thouless energy of about $E_{th}$ = \unit{100}{\micro \electronvolt} gives a fairly good fit of experimental data above \unit{150}{\milli \kelvin}. Below this temperature, the validity of Eq. \ref{ModelProc} breaks down as $k_B T < E_{th}/2\pi$. Such estimate of the  Thouless energy is comparable with the gap of aluminium ($e \Delta$ = \unit{130}{\micro \electronvolt}), therefore the junction lies in an intermediate  regime  between the long and short junction limit\cite{Universal}. 

The value of the Thouless energy allows us to estimate the coherence length at a temperature $T = \sfrac{E_{th}}{2 \pi k_B}$ representing the minimum $T$ where the relation \ref{ModelProc} holds. For this temperature ($T\sim$150~mK), using a  $v_F$ = \unit{4.2 \cdot 10^5}{m/s}\cite{SdH4,He2010} for the Fermi velocity of \BiSe\, we estimated a coherence length $\xi_n = \hbar v_F / E_{th}$ of \unit{\sim 3}{\micro \meter}, which is larger than the distance between electrodes (\unit{300}{\nano \meter}). This large value of $\xi_n$ is in agreement with the typical scale of the coherence length of traditional metals in S-N-S junctions\cite{clarke1969supercurrents,Clarke1971}. 

This gives a first hint of ballistic transport regime through the topologically protected surface states of the TI. The presence of defects with a spacing of few nanometers demonstrated by the STM topography images (see Fig.~\ref{STM1}), would not be consistent with a ballistic transport, without invoking the presence of topologically protected edge states.

We have collected $I_c$ values in a well distributed temperature range with respect to\citet{SdH1}, which allows a more accurate fitting of the $I_c(T)$.
In Fig.~\ref{Figure7}(b) we fit the data using a super current density  expression derived in the clean limit in Ref.~[\onlinecite{Galaktionov}], starting from  the Eilenberger equations, which is valid in our whole temperature range:
\begin{equation}
I = a \frac 2 \pi e k_F^2 \frac{k_B T}{h} \sin (\chi) \sum_{\omega_n} \int_0^1 \zeta \;d \zeta \frac{t^2}{Q^{1/2}(t,\chi,\zeta)} 
\label{eilen}
\end{equation}
where  $\zeta = k_x/k_F$  and $t  = D/(2-D)$ with D the transparency of the S-TI interface and $a$ is the cross section of the junction. Other parameters are reported in footnotes
\footnote{with $$ Q = \left[ t^2 \cos(\chi) + (1+(t^2 +1) \omega_n^2/\Delta^2 )\cosh\left(2 \omega_n L /\mu \hbar v_F \right) + \right.$$ $$\left. + 2 t \omega_n \Omega_n /\Delta^2 \sinh\left(2 \omega_n L/ \mu \hbar v_F\right) \right]^2 -(1-t^2)^2\Omega_n^2/\Delta^2 $$ where $\omega_n = 2 \pi k_BT(2n+1)$ are the Matsubara frequencies and $\Omega_n = \sqrt{\omega_n^2+\Delta^2}$. $\chi$ is the phase difference between the two superconducting islands.}.
The prefactor of Eq. \ref{eilen} is connected with the normal resistance of the sample $R_N$ which is unknown due to the shunt of the surface state with the bulk states. Therefore, following Ref.~[\onlinecite{SdH1}] we use it as a fitting parameter. The $I_cR_N$ product of the junction is lower than the values expected from the Ambegaokar - Baratoff model\cite{Barone}, which, on the other hand, is expected to hold only in the case of tunnel junctions. However the measured values of $I_cR_N$ are consistent with all other measurements in the literature\cite{SdH1, Morpurgo, StanfordTI} (see Table~\ref{table}).

The transparencies at the barriers are assumed to be equal to each other and they are given by $D\sim$~0.61, consistently from what is expected by the excess current of $I_{exc}/I_c \sim$~0.326. With these data we obtain a reasonably good fit in the whole temperature range. An attempt to fit experimental data in the diffusive regime through Usadel equations  does not provide a self consistent scenario, and yields unphysical values of the fitting parameters. The ballistic limit supports the hypothesis of  a transport through the topologically protected surface states of the TI.

If we naively assume that the measured normal resistance of the junctions is coming from the contribution of the 2D channel and calculate the corresponding $I_c R_N$ product, we obtain a value of \unit{1.57}{\micro e \volt} which is 2 order of magnitudes smaller than the bare gap of the Al electrodes. This is consistent with what is observed in literature as shown in Table~\ref{table}.  On the one hand, bulk shunt strongly increases the conductivity of the sample, thus reducing the measured $R_N$. The ratio between the surface and the bulk resistivity in our samples cannot be easily determined. Despite an almost identical geometry,  the samples  with a Pt buffer layer show a critical current which is an order of magnitude larger than those with Ti, reflecting different proximity in the buffer layer\cite{VortexNOI}.

\section{Summary and conclusions}

Exfoliated flakes of \BiSe\ have been characterized by longitudinal and transverse resistance (Hall effect) as a function of the external magnetic field. The presence of a 2D edge state with a carrier density of $n \simeq$~4.5~$10^{12}$~cm$^{-2}$ has been proved by the study of SdH oscillations, through the angle dependence of the Shubnikov-de Haas oscillations.

The existence of Dirac electrons in our samples has been  clearly demonstrated by STM measurements. Tunneling spectra acquired in applied magnetic fields perpendicular to the sample surface reveal the presence of Landau levels with unique characteristics of Dirac electrons. The most important feature is the n = 0 Landau level at the Dirac point, which is independent of the magnetic field.

The proximity effect in Al/\BiSe /Al Josephson junctions and SQUIDs was also studied. The samples have been characterized by studying the response of the critical current to an external magnetic field and in presence of microwave irradiation.  The temperature dependence of the critical current showed a ballistic transport, with a coherence length in the TI of about 3~$\mu$m. This coherence length is much longer than the typical distance between the scattering centers found by STM (of few nm), which give strong indication that the super-current is carried by the topological surface states.
The complete set of experimental data, assisted by a comparative analysis of different theoretical frameworks and the relative numerical codes, supports the notion of a super-current carried by the topological surface states. The analysis of the magnetic pattern and of the Shapiro steps as a function of the microwave power are consistent with a conventional CPR, and finally confirmed by phase sensitive experiments performed on reference SQUIDs.

These findings are not necessarily ruling out the possibility of observing Majorana fermions fingerprints, which could be washed out by the large number N of channels involved in the transport\cite{Snelder2013} (N of the order of 500). The single channel regime would be rather the ideal experimental configuration to detect MF. This would be realized by reducing the size of the flake to the nano-scale\cite{Potter2013}, thus reducing shunting effects of the conductive bulk. This work is in progress, and it is based on the `universality' of the ballistic regime, which occurs in \BiSe\-based barriers independently of the superconducting material of the electrodes.

We expect that new insights into the transport can also come from the study of switching current experiments, in analogy with recent works carried out on junctions with various barriers\cite{NOIPRL,NoiISCM,Bezryadin}. At the same time, by reducing the size of the flake and by engineering tunnel contacts,  one can get access to a new regime where charging effects are relevant\cite{Gustafsson}, possibly revealing new interesting physics, beyond the Majoranas.

\section{Acknowledgements}

The Authors acknowledge financial support by
MIUR-Italy through Prin project 2009 "Nanowire high critical temperature superconductor field-effect devices",
Progetto FIRB HybridNanoDev RBFR1236VV
COST Action - [MP1201] Nanoscale Superconductivity: "Novel Functionalities Through Optimized Confinement of Condensate and Fields" [NanoSC - COST] by the Swedish Research Council (VR) and SSF under the project Graphene based high frequency electronics and KAW under project "Dirac Materials".
One of us (S. C.) thanks the partial financial support of the Fonds Qu\'eb\'ecois de la Recherche Sur la Nature et les Technologies.
Work at Temple University (STM measurements) was supported by the U.S. Department of Energy, Office of Basic Energy Sciences, Division of Materials Sciences and Engineering under Award DE-SC0004556.

\bibliography{bibliography}

\end{document}